# Cognitive performance in open-plan office acoustic simulations: Effects of room acoustics and semantics but not spatial separation of sound sources

Manuj Yadav[1,a], Markus Georgi[2], Larissa Leist[3], Maria Klatte[3], Sabine J. Schlittmeier[2], and Janina Fels[1]

[1] *Institute for Hearing Technology and Acoustics, RWTH Aachen University, Germany*

[2] *Institute for Psychology, RWTH Aachen University, Germany*

[3] *Center for Cognitive and Developmental Psychology, RPTU Kaiserslautern-Landau, Germany*

The irrelevant sound effect (ISE) characterizes short-term memory performance impairment during irrelevant sounds relative to quiet. Irrelevant sound presentation in most laboratory-based ISE studies has been rather limited to represent complex scenarios including open-plan offices (OPOs) and not many studies have considered serial recall of heard information. This paper investigates ISE using an *auditory*-verbal serial recall task, wherein performance was evaluated for relevant factors for simulating OPO acoustics: the irrelevant sounds including speech semanticity, reproduction methods over headphones, and room acoustics. Results (Experiments 1 and 2) show that ISE was exhibited in most conditions with anechoic (irrelevant) nonspeech sounds with/without speech, but the effect was substantially higher with meaningful speech compared to foreign speech, suggesting a semantic effect. Performance differences in conditions with diotic and binaural reproductions were not statistically robust, suggesting limited role of spatial separation of sources. In Experiment 3, statistically robust ISE were exhibited for binaural room acoustic conditions with mid-frequency reverberation times, $T_{30}$ (s) = 0.4, 0.8, 1.1, suggesting cognitive impairment regardless of sound absorption representative of OPOs. Performance differences in $T_{30}$ = 0.4 s relative to $T_{30}$ = 0.8 and 1.1 s conditions were statistically robust, emphasizing the benefits for cognitive performance with increased sound absorption, which reinforces extant room acoustic design recommendations. Performance differences in $T_{30}$ = 0.8 s vs. 1.1 s were not statistically robust. Collectively, these results suggest that certain findings from ISE studies with idiosyncratic acoustics may not translate well to complex OPO acoustic environments.

---

[a] Also at: School of Architecture, Design and Planning, The University of Sydney, Australia. Electronic mail: manuj.yadav@akustik.rwth-aachen.de

## I. INTRODUCTION

Open-plan offices (OPOs) are notorious for the detrimental effect of their acoustic environments on occupants' activities [1–3]. Many such activities including proofreading, writing, etc. are believed to involve a critical role of short-term processing of verbal information in serial order [4]. Verbal short-term memory (STM) performance has been shown to be vulnerable to certain irrelevant sound sequences [5]. This phenomenon is referred to as the irrelevant sound effect (ISE) and has a rich history of development [6] since its first report [7]. ISE studies feature prominently in laboratory research with a focus on OPOs besides many OPO studies citing ISE research in general (summary in [8]). In fact, a model based on verbal STM serial recall performance [8] is integral to the room acoustic standard for OPOs [9].

However, a basic aspect shared by most ISE studies with or without an OPO focus remains largely unexplored. It involves acoustic presentation of the irrelevant sounds, and the corresponding ecological validity – 'the degree to which research findings reflect real-life hearing-related function, activity, or participation' [10] – of the ISE findings. The acoustic content in most ISE research has largely been idiosyncratic (e.g., spoken words, letters, etc.; details in I.A.2), with predominantly headphone-based diotic/monaural (same signal to both ears) sound reproduction (details in I.A.3), and an ambiguous role of room acoustics (details in I.A.4). Findings from such ISE studies represent low ecological validity for complex acoustic environments such as OPOs. Even for ISE studies using more representative OPO simulations [8,11–13], broader engagement with various germane aspects of acoustic presentation (e.g., irrelevant sounds, reproduction methods, room acoustic effects, etc.) has generally not been extensive. Moreover, the task used in most ISE studies involves serial recall of visually presented sequences (e.g., digits, words). Many situations or tasks in OPOs can involve processing aural information (e.g., from computers, headphones, telephones, neighbors) in the presence of irrelevant sounds. Hence, serial recall of aurally presented verbal information represents a valid consideration. Although studied less in comparison, serial recall of aurally presented information (e.g., [14–16]) has been shown to exhibit a similar effect as serial recall of visual information [15,16], which is a reasonable starting point for further ISE investigations in OPO simulations.

Hence, this study aims to investigate ISE using serial recall of aurally presented digits in laboratory settings wherein the acoustic presentation includes a closer representation of the mundane OPO speech and nonspeech acoustic environment, compared to extant studies. Herein, the acoustic presentation involves several considerations regarding the composition of the irrelevant sounds and the headphone-based reproduction method in laboratory settings.

### A. Considerations for irrelevant sound presentation in ISE studies

To set the stage, OPOs typically comprises of several spatially spread sources of intermittent (e.g., workstation sounds, doors, etc.) and continuous *nonspeech* sounds (e.g., heating, ventilation, and air conditioning (HVAC) noise, etc.), along with multi-talker *speech* from neighboring and far-away workstations, etc. [3]. Further, these sounds interact with the room acoustics in a complex manner [17]. To characterize such complex acoustic conditions in laboratory, arguably an appropriate level of sophistication in the simulation is needed. This is addressed in the following, starting with a more elaborate description of the ISE.

#### 1. *Duplex-mechanism account for ISE for serial recall*

For task-irrelevant background sound sequences to notably impair verbal serial recall performance, at least one of two conditions must be met [5]. The first refers to the *changing-state* characteristics of the sound sequence(s). Herein, to elicit the ISE , a certain degree of change in spectro-temporal state is required to impede verbal STM performance compared to *steady-state* sound sequences and/or quiet conditions [18]. This is attributed to the *interference-by-process* principle. In the current case, this principle posits that the order information for successively varying auditory-perceptive tokens in changing-state sound stream(s) is processed obligatorily and pre-attentively, which interferes with the processing of order required in the serial recall task [5]. The second condition refers to unexpected deviations in auditory-perceptive stream(s), known as the *attentional capture* mechanism [5,6]. Herein, the deviations exhibit attention grabbing potential (e.g., one's name being called, a brief tonal deviation, etc.) away from focal activity that may or may not require processing of serial order. The interference-by-process and attentional capture mechanism constitute the *duplex-mechanism* account for ISE for serial recall [5], which postulates functional differences between disruption due to these two mechanisms. Although, see [19] for the competing unitary account.

#### 2. *Content of irrelevant sounds in ISE studies and role of semanticity*

From an OPO perspective, speech and (sufficiently) changing-state nonspeech sounds that can be segmented into coherent auditory-perceptive stream(s) have been shown to exhibit ISE ([20] for office sounds; general review in [5]). This is relevant for OPO simulations, where nonspeech sounds and their typical regularity of occurrence can be deduced from audio recordings or literature [3]. The role of semanticity of speech, however, needs additional considerations. Within OPO occupants' surveys, which perhaps represent maximal ecological validity, intelligible speech, especially from nearby workstations [21], consistently ranks as the most disturbing component among all OPO sounds [1,2,21]. This is reflected in current OPO acoustic standards. These prioritize reducing speech transmission index (STI) across workstations as a key strategy for improving room acoustics [9] and overall acoustic comfort [22]. However, as per the duplex-mechanism account (section I.A.1), while meaningful speech can capture attention away from the focal task (e.g., due to emotional or personal relevance to speech, taboo words) [23–26], the role of meaningful speech is considered irrelevant for interference-by-process. This is because the serial recall task is presumed to not require extensive semantic processing ([27]; review in [6]). Yet, it is worth noting that speech has consistently been reported as the most potent distractor during serial recall [28–30]. Further, speech in participants' own language has generally been shown to be more disruptive in magnitude than speech in a foreign language, although not always significantly so [29].

Additionally, emerging evidence points towards automatic semantic processing of irrelevant sounds during serial recall. Such automatic processing can disrupt task performance when contextual expectations are not met. This includes categorical deviation (e.g., number in a word sequence) [31–33], or semantic mismatch (e.g., unexpected ending to a sentence) [19,25,34]. Unlike attentional capture, the categorical deviation and semantic mismatch effects (both arguably semantic effects) have been shown to be immune to top-down control. For instance, these effects have been observed to be immune to habituation over the course of experiment [25,26,33,34]. Further, these effects have been shown to occur for deviants that are not relevant for the participants [25,26,31]. Categorical deviation has additionally been shown to be immune to foreknowledge about the deviant [31,32] and being unrelated to working memory capacity [32]. These effects are hence being proposed as being functionally separate from the current formulations of attentional capture in serial recall [26,31,33]. Besides, physiological evidence from a pupillometry based study of auditory attention also suggests that, while performing a cognitive task, meaningful irrelevant speech in one's native language consumes more attentional resources than meaningless speech in a foreign language [35]. This indicates that although task-irrelevant, native speech may nevertheless be processed semantically, and perhaps be more attention grabbing than foreign speech. Determining the functional basis (if any), and potential differences between automatic semantic processing and attentional capture mechanism during serial recall, are beyond the scope of the current study. However, the higher magnitude of ISE for native vs. foreign speech, and the emergent categorical and semantic mismatch effects at least provide some basis for exploring semantic effects in simulated OPO environments.

Moreover, in most ISE studies (primarily with cognitive psychology roots), the irrelevant nonspeech content has traditionally included tones, music, etc., and/or spoken content including words, letters, repetitive sentences, etc. The spoken content is typically acoustically dry and clear speech, often spoken by a single talker, rather than in a natural or conversational speech style by multiple talkers (see [5,6,28,36] for summaries). Such relatively simple stimuli may be necessary/justified for certain investigations, e.g., for basic cognitive psychology, and indeed be ecologically valid for such purposes. However, they are restricted representations of everyday scenarios such as OPOs, which may involve aspects that are not possible to study using simplistic stimuli (e.g., fluctuating spatial location of speech, staggered storylines as in halfalogues over telephones, etc.). To address the role of more conversational speech to an extent, some ISE research with an OPO context has included studies with more complex stimuli. This includes using a recorded mix of speech and several nonspeech sounds common in offices ([20] included sounds from doors, computers, typing, telephones rings, etc.), using in-situ OPO sound recordings with speech and nonspeech sounds (e.g., [28,37,38]), and recordings of multi-talker speech with repetitive (e.g., [20,39,40]) or non-repetitive content [11,12] in mock office set-ups. However, no study (to the best of authors' knowledge) has considered both naturalistic and non-repetitive multi-source speech (in native *and* foreign language to the participants) *and* nonspeech sounds representative of the OPO context in a controlled setup (i.e., not uncontrolled in-situ recordings) using an *auditory*-verbal serial recall task.

### 3. *Acoustic reproduction format*

The *reproduction* of acoustic content involves additional spatial, binaural, and room acoustic considerations. Diotic/monaural (i.e., same signal to both ears) headphone-based reproduction implies spatially-fused 'in the head/internalized' perception of acoustic content for the participant, which oversimplifies binaural hearing in complex acoustic environments such as OPOs. However, diotic headphone-based reproduction of irrelevant background sounds is predominant in previous ISE studies. Exceptions include some studies using binaural (i.e., adequately considering and reproducing interaural differences) reproduction (e.g.,[11,13,17,17,39,41–43]).

Using binaural reproduction, externalization of sound sources can be achieved. This leads to more complex variations in the intelligibility/audibility of certain sources compared to diotic reproduction of the same stimuli (summary in [44]). Such changes to intelligibility/audibility in binaural reproduction can include contributions due to spatial release from *energetic* masking including binaural unmasking and better-ear listening [45]. Additionally, contribution from *informational* masking is also likely [46], at least for nearby talkers [47,48]. This is relevant if the unmasked sounds are salient enough to draw attention, which is of interest from an attentional capture perspective, and may have additional semantic relevance for naturalistic speech. In that regard, spatially separated multi-talker speech from loudspeakers in acoustically dry conditions has been shown to restore the capacity of irrelevant speech to reduce cognitive performance, compared to same multi-talker speech from spatially fixed condition [39]. However, the stimuli in [39] included repetitive speech, which underrepresents complex speech in OPO environments. A recent study compared visual-verbal serial recall performance in a baseline condition with steady-state ventilation noise condition, with diotic and binaural reproduction of irrelevant classroom sound conditions (nonspeech sounds mixed with foreign language multi-talker speech) [43]. Herein, the serial recall performance in the classroom sound conditions was significantly different than the baseline condition i.e., ISE was exhibited. Further, the performance difference between the diotic and binaural conditions did not vary significantly for either adult or child participants [43]. However, a comparison of traditional diotic reproduction and more acoustically accurate binaural reproduction of the same irrelevant acoustic content using an *auditory*-verbal serial recall task, and using perceptually salient speech and nonspeech sound mix representative of OPOs, has not been conducted yet. Such a comparison is a key consideration in the current paper.

### 4. *Room acoustics*

Beyond (anechoic) spatial separation in reproduction format, OPO room acoustics can further complicate characterizing impact of task-irrelevant background sounds on STM cognitive performance [49]. Serial recall performance during irrelevant speech with an unrealistic (at least for OPOs) reverberation time ($T$ in seconds) of 5 s was shown to be similar to performance in quiet [50]. Using a mix of OPO nonspeech sounds (telephones rings, door slams, ventilation, etc.) and conversations recorded anechoically in virtual settings, another study showed no significant difference between serial recall performance in $T$ = 0.7 vs. 0.9 s (several details including source placements unavailable) [51]. However, performance in these reverberant conditions was significantly more disruptive than in quiet (i.e., ISE was exhibited). Another study tested the effect of using a 3- or 15-voices mix (repetitive speech) originating 10 m from a simulated listener position for $T$ = 0.4 s and 1.0 s, and an anechoic condition [52]. The results showed that serial recall performance relative to the quiet condition reduced significantly in all conditions. These included conditions with increased reverberation time and/or number of voices (including an anechoic condition), except for the condition with 15 voices and $T$ = 1.0 s [52]. In the latter, the performance was not significantly different to the quiet condition, which the authors attributed to sufficiently reduced changing-state with multiple voices and high reverberation [52]. Moreover, the authors suggested long reverberation times as a possible solution to reducing speech-based distraction in multi-talker OPOs [52]. However, both the studies with more plausible OPO reverberation times [51,52] do not consider realistic spatial sound arrangements and background sounds. Most speech-based disruption in OPOs tends to be from intelligible speech from spatially-separated nearby workstations [1,21]. Hence, distant voices 'mixed/fused' (e.g., [52]) likely underestimate the changing-state characteristics of actual OPO speech. Besides, the current room acoustic perspective for OPOs [9] advocates reducing speech intelligibility between workstations with the combined use of sound masking and sound absorption. The latter leads to lowered $T$ values, which goes against the recommendation of high reverberation times in [52]. Yet, there is also a trend of eschewing sound absorption in many recent OPOs including those with activity-based working (ABW) [3]. From this perspective, investigating the effect of higher $T$ values is informative using realistic simulations of representative OPO room acoustics, which is considered in this paper.

Moreover, while not studied systematically in the current paper, early reflections (~ 50-100 ms after direct sound in a room impulse response) can assist with speech intelligibility [53,54], at least for nearby talkers. Hence, when implemented in room acoustic simulations (see section III), early reflections may potentially degrade cognitive performance and/or increase attentional focus when the reverberant decay is not too long [12]. This may need to be considered alongside the smearing effect due to reverberant energy. Overall, the role of realistic OPO room acoustics on ISE using representative background sounds is still largely unexplored.

### B. Motivations for the current paper

As expounded above, most ISE studies typically include a limited representation of complex acoustic conditions such as OPOs [55]. This paper investigates the role of several variables that are relevant for laboratory-based simulations of OPOs in ISE studies. Based on the literature review above, these variables include: *semanticity of speech* (native vs. foreign language for participants mixed with nonspeech sounds; section I.A.2), *spatial presentation of stimuli* (traditional diotic presentation vs. acoustically more accurate binaural reproduction of spatially spread sources; section I.A.3), and *room acoustics* (anechoic vs. representative OPO room acoustics; section I.A.4). Note that this is not an exhaustive list of relevant variables. It is indeed possible to further increase the complexity of the laboratory-based simulations, some of which will be addressed in section IV.

All variables except room acoustics are tested in two experiments (Experiments 1 and 2 below) that address anechoic presentation of irrelevant sounds in previous ISE studies. Experiment 3 focuses on room acoustic variations. Aural presentation of to-be-recalled sequence of digits is used in all experiments as a starting point in representing such information in OPOs, which is elaborated further in section II.A.3

Overall, the expectations are that while the ISE in serial recall may be exhibited in all conditions with changing-state sounds, a more realistic representation of OPO sound environment regarding spatial presentation of stimuli and room acoustics may contribute to the overall decline in cognitive performance. This overall decline may be more pronounced in conditions with meaningful speech than speech foreign to the participants. This is expected due to the speech content here being more salient, and thus potentially more attention grabbing compared to more rudimentary speech in traditional ISE studies, besides other semantic effects.

## II. EXPERIMENTS 1 AND 2

These experiments test the ISE in terms of two variables noted above for irrelevant sound stimuli that includes nonspeech sounds and speech: *the semanticity of speech*, and *spatial presentation of stimuli.* The main difference between the experiments was that the irrelevant speech was in either semantically meaningful native German (Experiment 1) or in semantically meaningless Hindi language (Experiment 2). Hence, effect of semanticity was tested in a more ecologically valid manner compared to studies using manipulations including spectral [56,57] or vocoding [58], etc. of speech. Such manipulations, while preserving temporal fluctuations, generally do so at the cost of sounding artificial. Each experiment had five acoustic conditions and irrelevant sounds were presented as continuous audio over headphones (Table 1). These conditions included *Quiet* as the baseline condition for cognitive performance. The other conditions include diotic and binaural reproductions of task-irrelevant nonspeech sounds only, and nonspeech sounds mixed with speech, respectively. All conditions (including *Quiet*) included low sound pressure level (SPL) noise representing HVAC noise in offices (see Table 1 and section II.A.1). The diotic conditions represent most traditional ISE studies where the same audio signal is presented in both headphone channels with generally no interaural difference and/or spatial cues. The binaural conditions in these experiments included spatialized sounds in anechoic conditions.

TABLE 1: Summary of the experimental conditions in Experiments 1 and 2 in terms of key acoustic and psychoacoustic metrics. SPLs reported are power sum of left and right ear $L_{A,eq,1min}$ values. *FDCC* (frequency domain correlation coefficient) and *FS* (fluctuation strength) further described in section II.A.5. *D*: diotic; *B*: binaural; NS: nonspeech; G: German; H: Hindi; HVAC: heating, ventilation, and air-conditioning.

| Experimental condition | Nonspeech sounds SPL (dB(A)) | Speech SPL (dB(A)) | Overall SPL (dB(A)) | *FDCC* | *FS* (vacil) |
|---|---|---|---|---|---|
| *Quiet*: HVAC noise only | - | - | 41.5 | 0.9 | 0.32 |
| Diotic nonspeech sounds ($D_{NS}$) | 55 | - | 55 | 0.5 | 1.0 |
| Diotic nonspeech sounds + speech ($D_{NS\_G}$ or $D_{NS\_H}$) | 52 | 52 | 55 | 0.5 | 0.9 |
| Binaural nonspeech sounds ($B_{NS}$) | 55 | - | 55 | 0.5 | 1.3 |
| Binaural nonspeech sounds + speech ($B_{NS\_G}$ or $B_{NS\_H}$) | 52 | 52 | 55 | 0.5 | 1.0 |

### A. Methods

#### 1. *Irrelevant OPO background sounds for both binaural and diotic conditions*

The goal here was to include sounds that are common in OPOs, which were selected based on actual office recordings (reported elsewhere [3]). First, pink noise shaped with a -5 dB/octave decay was used, which is representative of the **HVAC noise** spectrum in OPOs [3,59]. This noise was presented at $L_{A,eq,1min}$ (energy-equivalent A-weighted SPL over a 1-minute period) of 41.5 dB (Table 1). This represents relatively quiet/modern HVAC systems in OPOs [60], where the SPL is lower than what would be required for substantial sound masking [61]. The signal presentation was decorrelated between the two headphone channels for a more 'diffuse' noise perception. This provided seamless steady-state low SPL noise throughout the experiment, including during the *Quiet* condition (Table 1). The latter provides a better representation of 'quiet' conditions (i.e., without other non-HVAC sounds) in OPOs.

Second, the **nonspeech (NS) sounds** used included a diverse set of anechoic recordings of typical activities in OPOs. The overall choice of nonspeech sounds and how frequently each group of sounds occurred was based on listening to actual recordings of several medium-sized OPO. As such, some sounds were more frequent than others to depict activities in a typical OPO. Besides, it was ensured that each type of nonspeech sound was presented at irregular intervals to avoid repetitive patterns. The nonspeech sounds included those originating from relatively regular use of furniture, keyboard and mouse and other stationery items at workstations, and printer operation; and relatively irregular phone rings, footsteps, door opening and closing, and elevator bell (ping) [3]. The latter three sounds were especially irregular. In the *diotic* conditions, the sounds were presented without any spatial separation between source locations. In the *binaural* conditions, the sounds were spatially spread according to their source location in the simulated room. This spatial spread, combined with the randomized order of nonspeech sounds, resulted in spatial randomization (see section II.A.2).

Lastly, **multi-talker speech** was used. Herein, the goal was to present non-repeating multi-talker speech with rich content and where each talker represents one side of a telephone conversation (i.e., a halfalogue) at a workstation, which is a common scenario in OPOs [21,56]. Studies have shown that such halfalogues are not only perceived as being more annoying and distracting compared to hearing both sides of a dialogue [62,63], but also impede performance in cognitive tasks involving semantic processing [56,64,65]. For the speech recordings, four native German talkers (1 Female (F), 3 Male (M)) and two native Hindi talkers (1 M, 1 F) volunteered. To achieve gender parity in voices (i.e., 2 F and 2 M voices for each language), the voice of one of the male German talkers and half of the segments from each of the Hindi talkers were pitch shifted to sound like the opposite gender in post processing. Each recording lasted around 2 hours with breaks and included a pair of talkers having unscripted conversations on a variety of topics (e.g., vacations, current affairs, work, movies, etc.). The talkers per pair were familiar to each other, which further helped in eliciting naturalistic conversation balanced in terms of overall content per talker. Each recording (32-bit, 44.1 kHz sampling rate) was conducted in a hemi-anechoic room (hard floor) with talkers sitting on chairs facing each other at a 2 m distance. Each talker wore a DPA 4066 omnidirectional headset microphone positioned 7 cm from the center of lips, similar to previous research [12].

Further signal processing was done in MATLAB with the overall goal to create naturalistic halfalogues. Voiced segments per talkers were separated (2-50 s long). Per experimental condition, contiguous segments from four different voices (2 F) were arranged into 4-channel files where two talkers were simultaneously active at any time (see Fig 2. in [12] for a visual representation). Besides the natural pauses while speaking, a variable amount of silence (randomized between 0 – 4 s) was introduced between the segments. The long-term spectra of the talkers were matched to the 'normal' vocal effort spectra in [9] as used in previous studies [11,12]. Finally, the order of active talkers was algorithmically randomized. None of the conditions had any repetition in speech content and it was ensured that the conversation topics for halfalogues per condition were different.

For the *binaural* conditions, the randomly changing order of two simultaneous talkers meant that the spatial location of active talkers changed in tandem (i.e., randomly). Along with the randomized order of nonspeech sounds, this is an example of spatial changing state for the sound sources (i.e., talkers and nonspeech sounds). The effect of such spatial changing state was, however, not experimentally tested. In comparison, the perceived spatial location of active talkers and nonspeech sounds remained the same in the *diotic* conditions, although the order of active talkers and nonspeech sounds changed randomly.

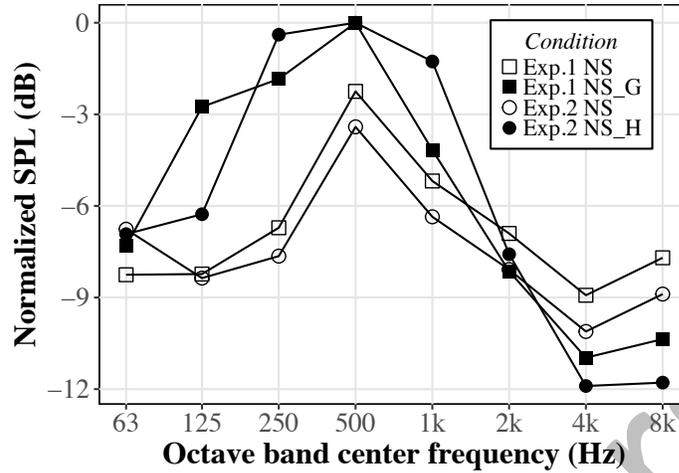

FIG. 1. Normalized SPL in the diotic nonspeech (NS) conditions and nonspeech and speech conditions in Experiments 1 (native German (G) speech) and Experiment 2 (foreign speech in Hindi (H)).

The overall presentation level of the conditions, except *Quiet*, was $L_{A,eq,1min}$ of 55 dB (Table 1), with a ~1.6 dB range ([54.0 dB, 55.6 dB], based on slow-weighted average of 15 s windows and 15 s hop size between windows), and $L_{A10}$ of 58.5 dB. The $L_{A,eq,1min}$ of the nonspeech sound conditions was matched across experiments as seen in Table 1. Figure 1 shows the spectra for the diotic conditions in Experiments 1 and 2, normalized with respect to the 500 Hz band (binaural conditions had similar spectra). Due to concentration of speech energy, the SPLs of conditions with speech mixed with nonspeech sounds (i.e., NS_G and NS_H) have substantially higher SPL compared to corresponding conditions with nonspeech sounds only (NS) in the 250 Hz and 500 Hz octave bands. There is also relatively higher SPL in the 125 and 1000 Hz in the NS_G and NS_H sound conditions compared to the nonspeech conditions, and somewhat relatively higher SPL in 4kHz and 8 kHz bands in the nonspeech conditions. Importantly, the spectral shape for NS_G and NS_H are relatively similar, i.e., limited spectral differences in the content of the two languages (German and Hindi).

In anechoic conditions with spatial separation, a SPL decrease for some sounds vs. the others will occur due to distance attenuation. This effectively means that the binaural sound conditions will have a lower SPL than the diotic conditions. However, to remove the influence of overall

SPL acting as a confounding factor, SPLs were matched across the diotic and binaural conditions, while maintaining the relative level of different sounds per condition (determined by the simulation output). The disadvantage is that the binaural sound conditions do not represent the true acoustics of modelled source distances (see section II.A.2). However, there is some evidence that the overall presentation SPL does not affect serial recall performance for SPLs typical of OPOs [12].

### *2. Simulated room and sound sources for binaural conditions*

For the *binaural* conditions, a room with workstations was modelled in Sketchup. The room has a rectangular floor plate and flat ceiling (267.3 m$^3$; 9 m (W) × 11 m (L) × 2.7 m (H)), and with one door and two windows (5.3 m × 2.1 m each) on separate walls. Twelve workstations were modelled over four straight rows (i.e., three per row) along the width of the floor plate. A mannequin sitting (anthropometric data from [66]) on a chair was modelled per workstation. Ear height was 1.2 m per mannequin. This model was imported within the geometrical room acoustics simulation software RAVEN [67] for auralization. All the surfaces were assigned an absorption coefficient of 1, i.e., no room reflections. The sitting mannequin modelled at the middle workstation in the last row was set as the sound receiver (listener) position. The listener was assigned head-related transfer function (HRTF) and headphone transfer function from the FABIAN database [68]. There were 16 sound sources in total. Twelve of these were assigned omnidirectional directivities to represent nonspeech sound sources and were placed at a height of 0.7 m each per workstation. Four sound sources were arranged at a (mouth) height of 1.1 m each at the corner workstations along two rows in front of the listener: two talkers at 3 m distance at ± 30 degrees, and two talkers at 5 m distance at ±20 degree. The latter four sound sources represented talkers (1 M, 1 F per row) and were assigned directivity of a singer (deemed sufficient here). The auralization output was 16 binaural room impulse responses (BRIR; one per source-receiver combination) for each of the experimental conditions except *Quiet* (Table 1). The BRIRs were convolved with the respective audio per sound source, which resulted in auralized files for 12 nonspeech and 4 speech sources per experimental condition.

### *3. Auditory stimuli for serial recall*

The to-be remembered digits were recorded as spoken digits (0-9) in clear speech in German by trained professional speakers using normal intonation and constant vocal effort [69]. The recordings were conducted in an anechoic chamber using a Neumann TLM170 microphone placed at a 1 m distance from the talkers. The voice used in this paper was the female voice B in [69], where more details of the recording are included. Each (anechoic) digit was 0.6 s in duration and presented diotically over headphones at 61 dB LA,Fmax in all conditions (Table 1). Hence, there was an effective signal-to-noise ratio (SNR) of at least 6 dB (20 dB for Quiet) for the difference between the to-be-remembered digits and the to-be-ignored background sounds (Table 1), which ensures sufficient intelligibility for the spoken digits. For comparison, SNR of 4 dB was sufficient in [16] for eliciting the ISE with auditory presentation of digits. Moreover, pilot experiments were conducted with three participants (neither the experiment participants nor the authors) whose data was not used any further. They reported that that the digits were clearly audible in all conditions, and were different sounding from the irrelevant sounds that included conversational (not clear) speech and nonspeech sounds. Further, it must be noted that the rather simple auditory presentation of digits was to establish a baseline and more complex presentations could be explored in future studies wherein elaborate investigations of masking mechanisms may be necessary.

*4. Calibration*

The auralized files with irrelevant sounds were mixed down to a 2-channel audio file per experimental condition (Table 1). The to-be-recalled digits were always presented diotically. The headphone (Sennheiser® HD650; Wedemark, Germany) output per channel (for each sound condition and each digit) was then calibrated using an artificial ear as described in [70]. The artificial ear used was a Brüel & Kjær (B&K, Nærum, Denmark) Artificial Ear Type 4153 with a B&K Type 4192 omnidirectional microphone capsule (conditioning through B&K Nexus Type 2690).

*5. Exploratory metrics quantifying changing-state characteristics*

While there is no consensus in terms of signal-based changing-state characterization, two metrics are included here for exploratory purposes: Fluctuation Strength (*FS*; [28,29,71,72]) and the Frequency Domain Correlation Coefficient (*FDCC*; [58,73,74]). *FS* (in *vacil*) is a psychoacoustic parameter that quantifies slower amplitude modulations in a signal (up to 20 Hz). *FS* was calculated using [75] as implemented within [76]. *FS* has been shown to be useful in characterizing ISE [28,72] in laboratory studies, although not always [29], and to exhibit a negative relationship with increasing number of workstations in actual OPOs [3]. *FDCC* aims to model changing-state characteristics (hence, the ISE) by analyzing spectral similarities between successive sound segments. *FDCC* has been used in studies with multi-talker speech and noise mix [58,77] and was calculated here based on [74]. *FDCC* values range from 0 – 1, with values approaching 1 exhibiting lower spectro-temporal changing state, and vice-versa for values approaching 0. While there are conceptual differences between these metrics [74], the general idea here is that decreasing *FS* and increasing *FDCC* relate to decreasing acoustic spectro-temporal changes in the irrelevant sounds.

However, both these parameters are currently limited in terms of reliably describing changing-state characteristics. Both *FS* and *FDCC* are monaural parameters and hence do not incorporate binaural effects. The thresholds for exhibiting an effect (i.e., just-noticeable differences (JNDs)) are not known for either of these parameters. Moreover, it is not clear whether these parameters sufficiently address all acoustic *and* perceptual aspects necessary to quantify changing-state characteristics [28,58].

Given these limitations, the interpretations of these parameters are largely considered tentative. *FS* and *FDCC* values for the experimental conditions (Table 1) expectedly show a negative relationship with each other. The baseline *Quiet* condition has the highest *FDCC* value (close to 1), and an appropriately low *FS* value, which can be used to predict the least changing-state characteristics and best performance within this condition. Table 1 shows that *FDCC* values (average of left and right ear) are quite similar across sound conditions except *Quiet*. Hence, the respective changing-state characteristics would be interpreted as more-or-less similar (again, JNDs are not known). However, *FS* values show more variation. Mixing speech with nonspeech sounds decreases *FS* values, which may represent increased energetic masking simply due to more sounds. Binaural conditions with spatially separate sources on the other hand can have release from masking for some sounds, which may explain the larger *FS* values (note again that *FS* is a diotic parameter). Overall, it is not clear the extent to which either of these parameters describe changing-state characteristics for the current conditions.

### *6. Experimental procedure*

Experiment 1 had 40 participants (22 Female; Age (years): *Mean* = 28.9, *SD* = 8.0) and Experiment 2 also had 40 participants (33 Female; Age (years): *Mean* = 21.5, *SD* = 6.9). None of the participants reported hearing impairment and had normal or corrected-to-normal vision. Most participants were native German speakers. Four participants in Experiment 1 were non-native German speakers who had 5-10 years of experience, which is considered sufficient here (performance results were similar to native German speakers). For both Experiments 1 and 2, the duration was approximately one hour per participant. This included time for instructions, practice, experimental tasks, breaks, and debriefing. Experiments 1 and 2 were conducted in separate hearing booths within the authors' respective institutes using identical audio equipment. This allowed basic assessment of the repeatability of the baseline conditions across two different laboratories. Both these hearing booths are acoustically treated, have low background noise (< 35 dB $L_{A,eq,1min}$), and included a desk with a computer screen and a keyboard for the participant. The participants remained seated and wore headphones throughout.

The experiment was conducted via a graphical user interface (GUI; coded in Python) on the computer screen. The GUI presented the experiment instructions, followed by the participants performing a practice run of three trials of the experimental task in *Quiet*. Each trial included performing the auditory-verbal serial recall task. This task included a set of eight digits as the to-be-remembered stimulus for recall, chosen without repetition from 0-9, and presented diotically as anechoic spoken digits in German (details in section II.A.3). Each digit was 0.6 s in duration with a 1.5 s interval between digits. Overall, the spoken digits roughly depicted a scenario of verbal information from a 'in the head' source (due to diotic reproduction). Note that due to the relatively high SNR (section II.A.3), the digits were heard clearly by the participants despite the irrelevant acoustic background. Hence, this does not represent a cocktail party scenario with generally adverse SNRs [44].

There were 12 trials per experimental condition, hence 60 trials overall (Table 1). Latin squares were used to balance the sequence of digits across trials. Participants were instructed to remember the digits in the order of presentation for subsequent recall, and to ignore the other sounds. Once all the eight digits were presented, there was a 6 s retention period. Following this period, the GUI presented three rows of digits (rows with 3, 2 and 3 digits), with each digit in black font in a square against a white background. The order of the digits across rows was randomized. The participants then recalled the digits by clicking corresponding digits in the GUI. Once a digit was selected, it was greyed out and could not be selected again. After each experimental condition, participants were notified via the GUI, and were allowed to take a short break. The procedure here and in Experiment 3 was approved by the Medical Ethics Committee at the RWTH Aachen University (EK 055-18)

## 7. Data analysis

The same statistical analysis method was used for Experiments 1 and 2, conducted using the software R (version 4.2.2). For each experimental condition per experiment, the first and last trial out of the 12 trials were removed. This was done to minimize any effect due to familiarization and fatigue during the first and last trials, respectively (result were similar with 12 trials). For the remaining 10 trials per experimental condition, a correct response was registered for every digit recalled in the exact serial order of its presentation (i.e., strict serial recall criterion). The total number of correct responses per condition were added, and percentage error per condition (*err%*) calculated. The relative difference in the error percentages between conditions (*RDerr%*) was also calculated based on [6].

The repeated-measures design of each experiment was incorporated as independently varying intercepts for the participants in corresponding Bayesian mixed-effects models. Each model included the experimental (sound) condition as the independent variable and *err%* as the dependent variable. Bayesian methods offer several advantages over frequentist (or classical) methods, particularly in terms of incorporating prior distributions and the interpretation of credible intervals. In Bayesian statistics, priors can be specified to reflect varying degree of prior knowledge (e.g., estimates of the ISE in literature) or beliefs about the parameters of interest. This minimizes the possibility of erroneous effect sizes that may result from fitting small or noisy data sets [78]. Additionally, given prior distributions and the data, Bayesian modelling provides the entire posterior probability distribution (PPD) per effect, rather than a point estimate of the most probable effect in the frequentist methods. Bayesian credible intervals (CIs), calculated here using the Highest Density Interval (HDI) of the PPD rather than quantiles, provide a more comprehensive picture of uncertainty. This is because Bayesian CIs indicate the range of values that have the highest probability of containing the true value of the effect [79]. In contrast, confidence intervals used in frequentist statistics only indicate the range of values that have a certain probability of containing the true parameter value, and do not provide information about the probability of the parameter being within that range. Finally, Bayesian methods also provide the ability to test for null effects by calculating the probability of a direction (PD), which is a tool for understanding the strength of evidence in support of a particular hypothesis (e.g., null hypothesis).

The Bayesian modelling was done using the *brms* (version 2.18) package [80] with mildly informative conservative priors. Prior distributions were sampled 80000 times: 8 independent chains of 11000 samples each, and discarding the first 1000 warm-up samples. No-U-Turn sampling was used [80] to avoid dependencies within each chain. More specifically, Gaussian distribution with a mean of 10 and a standard deviation of 5 was used as the prior to describe the difference in *err%* between *Quiet* and each of the other sound conditions (i.e., ISE per condition relative to *Quiet*). This prior is based on various studies where the ISE was demonstrated for mainly diotic presentation of nonspeech sounds with/without speech in conditions that represent office environments at least to some extent (e.g., [20,28]). The large standard deviation for the prior is chosen as a conservative representation for the variation in values amongst such ISE studies, and that no previous study has reported binaural reproduction effects for speech and nonspeech sounds mixed to the extent of the current study.

The calculated PPDs are summarized using median values and associated 95% CIs (calculated using HDIs). The latter is the interval with a 95% chance of containing the effect's true value, given the data and the model. Possible existence of an effect is inferred if the 95% CI does not contain zero when comparing two conditions, and PD > 97.5% in the effect's (positive or negative) direction. The latter provides a link to the frequentist framework since PD > 97.5% is mathematically equivalent to $p < .05$ [81]. CIs and PDs per PPDs were calculated using the *bayestestR* package [82].

## B. Results

Fig. 1 shows the modelled median errors (*err%*) in the serial recall task per experimental condition. Table 2 shows the effect sizes in the form of relative difference in error percentages

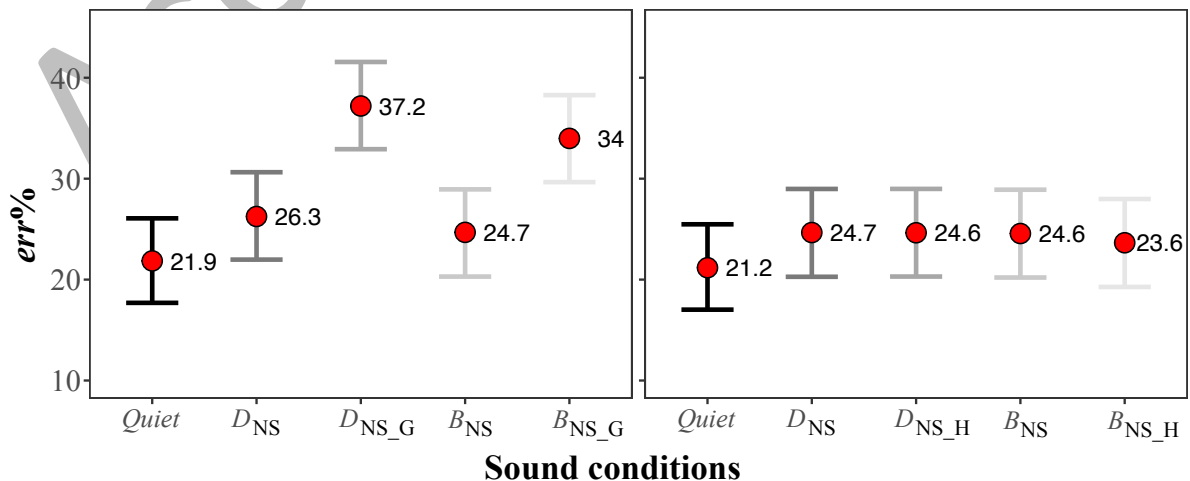

($RDerr\%$) for the experimental conditions in Experiments 1 and 2. Results per experiment are presented as follows.

FIG. 2. Median error percentages ($err\%$) and associated and 95% CIs for the posterior distribution modelled per condition for (left panel) Experiment 1 (native German (G) speech) and (right panel) Experiment 2 (foreign speech in Hindi (H)).

TABLE 2. Relative differences in median error percentages ($RDerr\%$) for conditions in rows 2 – 5 relative to the conditions in columns 2 – 5. The values (in %) are for Experiments 1 (German (G) speech) and 2 (Hindi (H) speech), respectively.

| Condition | Quiet | $D_{NS}$ | $D_{NS\_G/H}$ | $B_{NS}$ |
|---|---|---|---|---|
| $D_{NS}$ | 20.1, 16.5 | - | - | - |
| $D_{NS\_G/H}$ | 69.9, 16.0 | 41.4, -0.4 | - | - |
| $B_{NS}$ | 12.8, 16.0 | -6.1, -0.4 | -33.6, 0 | - |
| $B_{NS\_G/H}$ | 55.3, 11.3 | 29.3, -4.5 | -8.6, -4.1 | 37.7, -4.1 |

### 1. Experiment 1 (native German speech)

The $err\%$ were lower and statistically robust in the *Quiet* condition compared to the $D_{NS}$ (-4.4, 95% CI: [-8.0, -0.9]; 99.3% PD < 0), $D_{NS\_G}$ (-15.3, 95% CI: [-18.9, -11.8]; 100% PD < 0), and $B_{NS\_G}$ (-12.1, 95% CI: [-15.7, -8.6]; 100% PD < 0) conditions, i.e., ISE was exhibited. However, the difference in $err\%$ was not statistically robust between *Quiet* and the $B_{NS}$ (-2.8, 95% CI: [-6.4, 0.8]; 94.1% PD < 0) conditions.

The $D_{NS}$ condition had lower $err\%$ (statistically robust) than $D_{NS\_G}$ (-10.9, 95% CI: [-14.7, -7.0]; 100% PD < 0), and the $B_{NS}$ condition had lower $err\%$ than $B_{NS\_G}$ (-9.3, 95% CI: [-13.1, -5.39]; 100% PD < 0), i.e., meaningful speech increased the $err\%$ relative to nonspeech sounds. However, the $err\%$ differences between $D_{NS}$ and $B_{NS}$ (1.6, 95% CI: [-2.4, 5.4]; 79.1% PD > 0), and $D_{NS\_G}$ and $B_{NS\_G}$ (3.2, 95% CI: [-0.7, 7.1]; 94.6% PD > 0) conditions were not statistically robust, i.e., no robust changes due to the more realistic binaural reproduction.

### 2. Experiment 2 (foreign speech in Hindi)

The $err\%$ were lower and statistically robust in the *Quiet* condition compared to the $D_{NS}$ (-3.5, 95% CI: [-6.8, -0.1]; 98% PD < 0), $D_{NS\_H}$ (-3.5, 95% CI: [-6.8, -0.1]; 98% PD < 0), and $B_{NS}$ (-3.4, 95% CI: [-6.7, -0.1]; 97.8% PD < 0) conditions, , i.e., ISE was exhibited. However, the difference in *err*% was not statistically robust between the *Quiet* and $B_{NS\_H}$ (-2.5, 95% CI: [-5.8, 0.9]; 92.7% PD < 0) conditions.

With speech that was not meaningful to the participants, the *err*% differences were not statistically robust between the nonspeech and foreign speech conditions, i.e., between $D_{NS}$ and $D_{NS\_H}$ (0.02, 95% CI: [-3.5, 3.6]; 50.4% PD > 0), and between $B_{NS}$ and $B_{NS\_H}$ (0.9, 95% CI: [-2.6, 4.5]; 69.8% PD > 0). Similar to Experiment 1, the *err*% differences between the diotic and more realistic binaural conditions, i.e., between $D_{NS}$ and $B_{NS}$ (0.1, 95% CI: [-3.4, 3.6]; 51.9% PD > 0), and between $D_{NS\_G}$ and $B_{NS\_G}$ (1.0, 95% CI: [-2.5, 4.6]; 71.3% PD > 0) conditions, were not statistically robust.

### C. Discussion for Experiments 1 and 2

As expected, and in line with most previous research, ISE was exhibited for the diotic conditions in both experiments, albeit using an auditory-verbal serial recall task here. However, the main finding is the extensive difference in the conditions with irrelevant speech across the two experiments (Fig. 2). In Experiment 1, ISE was substantially larger in both diotic and binaural conditions with meaningful speech (i.e., $D_{NS\_G}$ and $B_{NS\_G}$), compared to Experiment 2 (i.e., $D_{NS\_H}$ and $B_{NS\_H}$) which included foreign speech that was meaningless to the participants. This difference is evident from the fact that speech conditions in Experiment 2 had 49% higher *RDerr*% (averaged across diotic and binaural conditions; Table 2).

An important consideration here is whether these results can be explained using changing-state variations, at least based on the (exploratory) signal-based (i.e., non-semantic) parameters in section II.A.5 and Table 1. *FDCC* and *FS* predicted ISE in all sound conditions, if one assumes sufficient variation in values compared to the *Quiet* condition, which is mostly consistent with the results. However, neither *FDCC* nor *FS* is designed to predict an effect based on the language of speech used. Moreover, more disruption was expected in binaural conditions due to spatial release (hence, potentially increased changing-state characteristics) of sound sources compared to spatially fused sounds in diotic conditions. Indeed, this could be predicted from corresponding higher *FS* values in binaural conditions (note that neither *FS* nor *FDCC* is a binaural parameter). However, ISE in binaural conditions was not statistically different from the diotic conditions, which is inconsistent with the corresponding *FS* prediction. Therefore, the finding of higher *err*% in native vs. foreign speech conditions is hard to explain using signal-based spectro-temporal/changing-state considerations. Instead, pending future investigations, these findings suggest a semantic effect wherein increased attentional capture during serial recall can be expected for semantically meaningful native speech. Additionally, possible contribution are likely due to automatic semantic processing during serial recall that is proposed in some recent studies [25,26,31,33]. The role of such factors in ISE for native vs. foreign speech conditions should be explored in future studies, including effects due to intonation, etc.

Note however that the serial recall performance in the binaural conditions relative to *Quiet* were equivocal across the two experiments. Firstly, ISE was exhibited in the $B_{NS}$ condition in Experiment 2 (and barely so) but not in Experiment 1. The reason for this discrepancy is not clear except for potential interlaboratory differences despite the matching audio reproduction. From a signal perspective, mixing speech with nonspeech sounds may reduce spectro-temporal variation due to increased energetic masking. This is consistent with the *FS* value being higher in the nonspeech conditions compared to conditions with speech, while *FDCC* was the same across conditions (Table 1). In fact, *FS* value for $B_{NS}$ was largest of all the conditions, yet it showed one of the lowest *err%* in both experiments. Hence, in this case, *FS* is incorrect in predicting ISE and higher disruption in $B_{NS}$. Moreover, while *FDCC* did not suggest higher disruption in $B_{NS}$, it also predicted ISE, and is partially incorrect.

The second discrepancy in the binaural conditions across experiments was that while statistically robust ISE was exhibited for *Quiet* vs. $B_{NS\_G}$ in Experiment 1 (notwithstanding the semantic effect discussed above), no ISE was exhibited for *Quiet* vs. $B_{NS\_H}$ in Experiment 2. The latter is unexpected since previous research generally shows statistically significant reduction in serial recall performance in foreign speech compared to quiet and/or steady-state background noise. While most such studies (e.g., [29]) include diotic reproduction, a closer comparison is possible with a recent study which included *Quiet*, diotic and binaural conditions that resembled those in Experiment 2 [43]. The main difference was that this previous study included classroom nonspeech sounds mixed with multi-talker speech of children in Hindi [43] (recorded using the same method as in II.A.1), which was a language foreign to the participants. For the adult participants in that study [43], ISE relative to *Quiet* was reported in the diotic condition (*RDerr%* ≈ 21%) like that for $D_{NS\_H}$ in Experiment 2 (*RDerr%* = 16%; Table 2). However, ISE relative to *Quiet* was also reported in the binaural condition (*RDerr%* ≈ 29%) [43] unlike that for $B_{NS\_H}$ in Experiment 2 (*RDerr%* ≈ 11%). While the *RDerr%* values for the diotic conditions in Experiment 2 and in [43] are somewhat similar, those for the binaural conditions are not. To address the latter, the role of some methodological differences between Experiment 2 and [43] must be noted. Primarily, this included differences in the speech (adult's voices here vs. children's voices in [43]) and the nonspeech sounds used. Moreover, the much lower sound reproduction SPL in Experiment 2 (< ~ 8 dB than in [43]) that may be relatively less disruptive [83]. To consider a signal-based assessments of the audio used, while the *FDCC* values for $D_{NS\_H}$ and $B_{NS\_H}$ in Experiment 2 were identical (Table 2), the *FDCC* values for the diotic and binaural conditions in the previous study [43] changed from 0.7 to 0.6 (although with identical *FS* of 1.3). This implies larger differences in the spectro-temporal/changing-state characteristics in the diotic and binaural conditions in the previous study [43]. These larger differences may explain the ISE in both conditions compared to *Quiet* in [43], but not in the binaural condition in Experiment 2. Besides issues in predicting ISE in $B_{NS}$ noted above, the role of *FS* seems ambiguous in this case as well, as it did not vary much between the diotic and binaural conditions in the previous study [43]. While the *FS* values varied in Experiment 2 here, the predictions do not match the results as discussed above. Hence, the different audio used in Experiment 2 and a previous study [43], and correspondingly different changing-state characteristics (based on *FDCC* values) may explain the different results regarding ISE in *Quiet* vs. $B_{NS\_H}$. However, these considerations do not explain why statistically robust ISE was not exhibited at all in $B_{NS\_H}$ in Experiment 2. Exploring this further is proposed for future studies wherein the role of binaural reproduction and changing-state characteristics may be tested in more details. This may include using systematic variation of *FDCC* values as *FS* values did not seem useful in the current context.

To continue with sound reproduction, the role of energetic masking (including its relationship with changing-state characteristics) between the diotic and binaural conditions was considered in the previous sentences. Furthermore, the speech content (two talkers randomly active at any time from out of four locations) may have involved some variation in information masking across these conditions. However, the role of informational masking is harder to assess here based on previous findings. This is because the closer talkers in Experiments 1 and 2 are at longer distances and wider angles than talkers in [48], where informational masking with close talkers was found. Hence, a detailed examination of masking principles is not possible here. Yet, the results overall suggest that differences in cognitive performance in the more ecologically valid binaural multi-talker reproduction of meaningful speech (in German, and without room acoustics; see Experiment 3) mixed with OPO nonspeech sounds, relative to corresponding monaural reproduction (common in traditional ISE studies), were not statistically robust.

Finally, to investigate the role of auditory vs. visual presentation of digits for serial recall, it is worth comparing the *RDerr*% values in current vs. previous studies. For instance, *RDerr*% of around 36% using a visual-verbal serial recall task was reported for diotic speech (meaningful to participants, in English) mixed with office nonspeech sounds (similar sounds as the current study) relative to a condition with just office nonspeech sounds in [20]. In comparison, the *RDerr*% in $D_{NS\_G}$ relative to $D_{NS}$ is around 41% (Table 2) using the auditory-verbal serial recall task. The 5% difference in *RDerr*% for relatively similar sound conditions between the current study and [20] is arguably minor in this context. Besides, the *RDerr*% values for $D_{NS}$ vs. *Quiet* comparisons in Experiments 1 and 2 (20% and 17%; Table 2) are similar to several previous studies that have included similar comparisons. For instance, median *RDerr*% of around 26% and a range of -5% to 62.7% has been reported in [6]. This included studies with comparisons between various types of diotic (non-office) nonspeech sound conditions (similar in principle to $D_{NS}$) and quiet [6]. Moreover, as discussed previously, the *RDerr*% values in Experiment 2 and in a study with a similar design of diotic audio [43], albeit using a visual-verbal serial recall task, are quite close. These *RDerr*% comparisons between Experiments 1 and 2 and previous studies, especially those with meaningful speech conditions (e.g., [20,43]), suggest that current results with an auditory-verbal serial recall task are at least plausible, if not expected to be similar, compared to visual-verbal serial recall tasks. Note again that similar results for visual-and auditory-verbal serial recall task have been shown previously for irrelevant music, and irrelevant (meaningful) speech conditions, relative to quiet [16]. Yet, foreign speech in the binaural condition in Experiment 2 did not lead to robust increase in *err*% values relative to *Quiet* or binaural nonspeech OPO sounds. Hence, it cannot be ruled out entirely that the auditory presentation of digits in native language did not influence the results. Moreover, while the presentation level of digits was determined here based on pre-experiment tests (section II.A.3), the effect of masking mechanisms may be relevant for more complex presentations. These aspects regarding the digit presentation, however, were not tested directly and are recommended as future research using variable SNR and more detailed investigations of potential masking mechanisms .

To summarize, ISE was exhibited in irrelevant nonspeech conditions and conditions with native (hence, meaningful) speech. Moreover, the reported STM performance in anechoic OPO simulations using speech in native language for the participants was much worse than performance in foreign speech, suggesting a semantic effect beyond changing-state considerations. Detailed explorations of this semantic effect are recommended in future studies. ISE predictions based on *FDCC* values seemed more consistent with the results compared to *FS* values, although neither parameters could account for all the findings here. This underlines their current state of development. This includes these parameters being approximations of changing-state characterization and not aiming to account for either attentional capture or semantic effects. Furthermore, the results here suggest limited benefit of binaural reproduction in anechoic OPO simulations. However, since the current binaural reproduction used generic HRTFs, the effect of more individualized or individual HRTFs, at least in terms of attention-grabbing aspect of sounds, remains a valid question for future research. Besides, since both experiments included anechoic conditions, a logical further step includes testing the effect of room acoustic conditions that are representative of OPOs on STM performance, which is studied in the next experiment.

## III.    EXPERIMENT 3

This experiment investigates the role of *representative room acoustics* in binaural OPO simulations using typical OPO nonspeech sounds, and multi-talker speech in participants' native language (German). The chosen room acoustic scenarios include reverberation times ($T_{30}$ in seconds) spanning a mid-frequency (average of 500 Hz – 2 kHz octave-band center frequencies) $T_{30}$ range of 0.4 s – 1.1 s typical of most OPOs [3]. This $T_{30}$ range is moreover comparable with similar (presumably broadband) *T* range of 0.4 s – 1 s in previous studies [51,52]. Besides the reverberation time, the ratio of early and late energy was considered by keeping the clarity index ($C_{50}$ in dB) within an acceptable range.

### A. Methods

#### *1. Simulated room acoustics and irrelevant sounds*

The same room setup as described in section II.A.2 is used here. However, the room surfaces and furnishings are now assigned typical absorption and scattering coefficients within RAVEN [67]. The various material groups in the room included ceiling tiles, floor (with carpet), chairs with sitting persons, tables, windows, door, etc., which provided a room averaged mid-frequency $T_{30}$ of 0.4 s and $C_{50}$ of 13.1 dB (RA-1 in Table 3). These values represent 'high' speech clarity in relatively 'high' sound absorption conditions, also shown in the STI values representing 'good-excellent' speech intelligibility for single talker-receiver scenarios [84]. Note that the STI values here are strictly for illustrating the room acoustics variations, as STI is not well defined for fluctuating background noise (section II.A.4).

Starting from RA-1, absorption coefficients of the materials were iteratively adjusted using the MATLAB-based ITA-Toolbox [85] to derive the conditions RA-2 and RA-3 in Table 3. This was done while ensuring that the spectral shape of reverberation ($T_{30}$ plotted for octave-band center frequencies) remained similar across conditions RA-1 to RA-3 (3 – 5 in Table 3) over the mid frequencies. During this process, absurd absorption coefficients for surfaces were also avoided. RA-2 could nominally be ascribed relatively 'medium-quality' room acoustics for speech communication (high $T_{30}$ and fair-good STI and $C_{50}$) for its room volume. However, RA-3 is edging towards relatively 'bad' room acoustics for intelligible speech communication with relatively high $T_{30}$ and poor STI and $C_{50}$, achieved by mostly reflective surfaces. Moreover, the potentially excess reverberant energy in RA-3 (and perhaps RA-2) leads to even higher SPL of background speech. Such high SPL background speech potentially provides some beneficial speech masking and 'live' ambience preferred in some workplaces. Conversely, such high SPL background speech could also be perceptually more annoying and distracting than quieter background noise with lesser reverberant energy buildup [3,21]. However, exploring such subjective aspects is outside the current scope of research. Besides, the nominal assessment here could be questioned within certain contexts (see Figure 8 and linked discussion in [3]). Yet, RA-1 to RA-3 at least represent OPO conditions derived using acoustically more appropriate simulation methods than previous studies [51,52].

The binaural room impulse responses (BRIRs) from the procedure above were convolved with the same nonspeech sounds as in section II.A.1. The multi-talker speech used had different content from the same talkers. The calibration procedure and the changing-state metrics are as described in section II.A.3 and II.A.4, respectively. The SPL in all the room acoustic conditions was kept the same (section II.A.2) despite the tendency of reverberant energy to increase SPLs with higher reverberation times. However, the spectro-temporal smearing was preserved. As in Experiments 1 and 2 (Table 1), *FDCC* values were similar in magnitude for the reverberant conditions (3 – 5 in Table 3), while *FS* values changed slightly over the same conditions.

### 2. Experimental procedure

Experiment 3 had 40 native German speaking participants (23 Female; Age (years): $M = 25.3$, $SD = 6.4$). The rest of the experiment details are the same as in section II.A.6., including experimental duration, location (same as Experiment 1), diotic reproduction of auditory digits for serial recall task, etc. The only difference was the experimental conditions used. These included the room acoustic conditions (Table 3), besides also including the anechoic $B_{NS\_G}$ condition (referred to as *Anechoic* in the following) from Experiment 1 and the *Quiet* condition (Table 1).

TABLE 3. Experimental conditions for Experiment 3. Conditions 1 and 2 are the same as conditions 1 and 5 ($B_{NS\_G}$) in Table 1. SPLs are power-averaged $L_{A,eq,1\,min}$ values for left and right ears. Mid-frequency $T_{30}$ (reverberation time) and $C_{50}$ (clarity index) presented are room averaged values. STI values presented as the average and range (in brackets) over each talker-receiver configuration. *FDCC* (frequency domain correlation coefficient) and *FS* (fluctuation strength) further described in section II.A.5.

| Experimental condition | SPL (dB(A)) | $T_{30}$ (s) | $C_{50}$ (dB) | STI | *FDCC* | *FS* (vacil) |
|---|---|---|---|---|---|---|
| 1. *Quiet*: HVAC noise only | 41.5 | - | - | - | 0.94 | 0.32 |
| 2. *Anechoic* | 55 | - | - | ~ 1.0 | 0.51 | 1.02 |
| 3. RA-1 | 55 | 0.4 | 13.1 | 0.73 (0.70-0.82) | 0.75 | 0.88 |
| 4. RA-2 | 55 | 0.7 | 5.3 | 0.64 (0.60-0.73) | 0.77 | 0.71 |
| 5. RA-3 | 55 | 1.1 | 2.9 | 0.57 (0.53-0.65) | 0.81 | 0.58 |

### B. Results

Fig. 3 shows the percentage of incorrect recalls (*err*%), and Table 4 shows the *RDerr*% for comparisons between experimental conditions. The *err*% differences between *Quiet* condition and each of the other experimental conditions were statistically robust, which included the *Anechoic* (-12.3, 95% CI: [-15.5, -9.1]; 100% PD < 0), RA-1 (-10.6, 95% CI: [-13.9, -7.5]; 100% PD > 0), RA-2 (-6.5, 95% CI: [-9.7, -3.4]; 100% PD < 0), and RA-3 (-6.5, 95% CI: [-9.7, -3.4]; 100% PD < 0) conditions. The *RDerr*% values for comparisons between experimental conditions shared between Experiment 1 (*Quiet* vs. $B_{NS\_G}$; Table 2) and Experiment 3 (*Quiet* vs. *Anechoic;* Table 4) were fairly similar.

Compared to the *Anechoic* condition, the *err%* differences were lower in the room acoustics conditions and were statistically robust relative to RA-2 and RA-3 (5.8, 95% CI: [2.4, 9.2]; 100% PD > 0 for each condition) but not relative to RA-1(1.6, 95% CI: [-1.78, 5.1]; 83% PD > 0). The *err%* differences were however statistically robust between RA-1 and both RA-2 (4.1, 95% CI: [0.7, 7.5]; 99.2% PD > 0) and RA-3 (4.1, 95% CI: [0.8, 7.6]; 99.1% PD > 0). Finally, the *err%* difference between RA-2 and RA-3 was not statistically robust (0.03, 95% CI: [-3.3, 3.5]; 51% PD > 0).

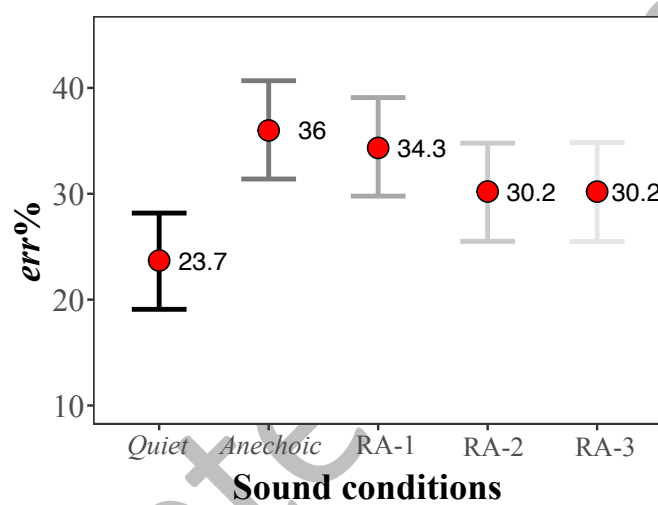

FIG. 3. Median error percentages (*err%*) and 95% CI for the posterior distribution modelled per condition for Experiments 3 with German speech and OPO nonspeech sounds presented binaurally (see Table 3).

TABLE IV. Differences in median error percentages for conditions in rows 2 – 5 relative to the conditions in columns 2 – 5 (i.e., *RDerr%*).

| Condition | *Quiet* | *Anechoic* | RA-1 | RA-2 |
|---|---|---|---|---|
| *Anechoic* | 51.8 | - | - | - |
| RA-1 | 44.9 | -4.6 | - | - |
| RA-2 | 27.4 | -16.1 | -12.0 | - |
| RA-3 | 27.5 | -16.0 | -12.0 | 0.03 |

### C. Discussion for Experiment 3

This experiment shows that serial recall performance in the *Quiet* condition was better (statistically robust differences) than other conditions including *Anechoic*, and room acoustic conditions with representative $T_{30}$ values for medium sized OPOs ($C_{50}$ variation in Table 3), and hence, exhibited the ISE. The relatively similar *err%* in the *Anechoic* and RA-1 (Table 3) conditions, which were not statistically robust, imply that a 'moderately-highly' absorptive room with high speech clarity may not be optimal for STM performance. This is consistent with the current recommendations in room acoustic design where speech clarity/intelligibility reduction between OPO workstations is the key consideration [9].

Moreover, [52] had reported ISE in a condition with $T = 0.4$ s in an OPO simulation with spatially fused multi-talker speech. The current results regarding RA-1 are hence consistent in principle with [52] in terms of ISE but not in terms of *RDerr%*. The *RDerr%* values in [52] between the quiet and $T = 0.4$ s condition were 9% and 4% for 3 and 15 voice mixes, respectively. These values, which are much lower than the *RDerr%* of 44.9% for *Quiet* vs. RA-1 (Table 4). While *err%* decreased with decreased sound absorption in RA-2 and RA-3 relative to RA-1, also consistent with [52] at least in terms of the direction of the effect, *err%* differences between RA-2 and RA-3 were not statistically robust. The latter is consistent in principle with [51] wherein serial recall performance in $T = 0.7$ s vs. 0.9 s conditions did not vary significantly. However, *err%* in RA-3 was still higher compared to *Quiet* (difference statistically robust). This is not consistent in principle with the main finding in [52]. Therein, serial recall performance was not significantly different in quiet compared to a condition with $T = 1.0$ s and with irrelevant speech from 15 voices at a 10 m distance [52]. The *RDerr%* were 1% and 7% higher in 3-and 15-voice mixes, respectively, with $T = 1.0$ in [52] and are quite different from 27.5% for RA-3 vs. *Quiet* (Table 4).

Although both refs.[51,52] used a range of $T$ similar to the current $T_{30}$, they have a rather limited and ambiguous representation of room acoustics and speech sources in OPO simulations. This likely leads to different conclusions and *RDerr%* between conditions across studies. Further, besides not including any OPO nonspeech sounds, the speech in [52] did not include nearby talkers, whose speech is generally intelligible and considered the primary source of auditory distraction in OPOs [1,2,21]. Hence, the current results can be considered a more ecologically valid comparison between different room acoustic conditions representation of OPO scenarios. As such, the current findings disagree with the (rather unintuitive) suggestion in [52] that increased $T_{30}$ (around 1.0 s) with distant and spatially-fused multi-talker 'babble' can exhibit serial recall performance similar to quiet conditions in OPOs.

Moreover, increased $T_{30}$ (i.e., reduced sound absorption) values for a fixed volume come with their own set of problems. These include increased overall reverberant SPL, etc., and designers generally tend to control the reverberant energy in OPOs. Note that this discussion does not intend to ascribe a special role to $T_{30}$, which would lead to a rather simplistic (even misinformed) assessment of room acoustics at best. As outlined in several studies [1,9,21,86,87], a careful *combination* of various room acoustic criteria is needed to achieve 'good' acoustics in OPOs. One such aspect of sensible room acoustic design that was not included in the current simulation was the use of partitions/screens between workstations, which was partly to allow comparisons with simulations in refs.[51,52]. Sound absorptive screens/partitions are typically used in OPO room acoustic design to allow better attenuation of direct sound between workstations, implying more effective reduction of speech clarity/intelligibility. In comparison, a previous study with almost the same multi-talker speech simulation (in English, and without nonspeech sounds) as the current had included absorptive screens between workstations. This previous study had reported *RDerr%* of 40% for a condition with $T_{30}$ = 0.5 s relative to quiet [12], which is relatively similar to *RDerr%* of 45% for RA-1 vs. *Quiet* in the current study (Table 4). This comparison underscores again the general difficulty in sufficiently reducing speech intelligibility between the nearest workstations using sound absorption alone, and where additional sound masking may be beneficial [1,21].

From another (and topical) perspective, the situation in RA-2 and especially RA-3 depicts recent trends in OPOs where screen/partitions are eschewed including within offices supporting activity-based working [3]. The current results do not support such practices without additional considerations such as sound masking, work type, etc. [22]. This is because the current findings show that STM performance in conditions with/without high sound absorption but without screens/partitions may still exhibit substantial ISE. However, there are a few caveats that need to be considered. Lombard speech [88], wherein speech with higher vocal effort is likely in reverberant conditions [89], was not implemented here. While there is some evidence that the overall SPL within OPOs may not be a major factor mediating ISE [12,90,91], this needs to be tested in simulations using Lombard speech corresponding to varying room acoustics. This would allow more realistic comparisons between different room acoustic conditions. In RA-2 and RA-3, the task performance was almost the same (and around 12% better than RA-1). Although higher $T_{30}$ along with lower $C_{50}$ than RA-3 may be achievable, it would perhaps be unreasonable for an OPO of current volume. In that regard, one may consider RA-2 as a compromise, with *err%* in serial recall performance similar to RA-3 but lower (statistically robust) than RA-1. However, the 'sweet spot' for OPO room acoustics is a much broader topic (see [9]), which needs to be tested using serial recall and other OPO tasks (e.g., writing, etc.) with variable room acoustics.

Consequently, the results here only represent the role of sound absorption without screens/partitions in OPO room acoustic design in terms of performance in STM tasks. While the current results represent higher ecological validity than previous ISE studies, many areas of improvements are possible. In this regard, as discussed in section II.C, the effect of better source localizations with individual/individualized HRTFs and headtracking needs to be tested. The to-be-recalled stimuli were presented as acoustically 'dry' spoken digits, i.e., did not change according to room acoustics. This unrealistic aspect of experimental design could be considered in future studies. Moreover, the current experiment is limited to medium-sized OPOs, with two nearby talkers and two talkers that are further away, along with representative OPO nonspeech sounds that are spatially distributed. Larger OPOs will have a more complicated assortment of speech and nonspeech sound sources, and the spatial release from masking can be quite complicated. Hence, the generalization of the current results needs caution beyond the conditions considered and more systematic studies to explore the effect of nearby sources.

In terms of the metrics to explore changing-state characteristics, the values are more informative compared to Experiments 1 and 2 (Table 1). The values of both *FDCC* and *FS* represent the *err%* across conditions except for RA-2 vs RA-3. One major difference in this experiment was that the irrelevant sound content remained the same, whereas in Experiments 1 and 2, the content changed across conditions. Hence, it seems likely that *FDCC* and *FS* may be more useful in some cases than others, and the role of these metrics needs elaboration in future studies.

## IV. GENERAL DISCUSSION AND OUTLOOK

The overall aim of this paper was to investigate the well-established ISE paradigm for STM serial recall using representative acoustics for OPO environments. To summarize, in Experiments 1 and 2, ISE was exhibited as expected in all but one condition with OPO nonspeech sounds. Further, ISE was exhibited in all conditions with meaningful speech mixed with nonspeech sounds in Experiment 1, and in the diotic condition with foreign speech mixed with nonspeech sounds in Experiment 2. However, the magnitude of the ISE was much larger in conditions with native speech relative to foreign speech, which suggests a semantic effect. Moreover, findings from Experiments 1 and 2 show that STM cognitive performance may not vary between *anechoic* diotic and binaural reproduction, suggesting limited role (yet no disadvantage) of incorporating realistic spatial separation and interaural cues in the irrelevant stimuli. Experiment 3 showed that with realistic room acoustics for OPOs, serial recall task performance may not approach performance in *Quiet* as suggested in previous studies (e.g., [52]). As such, findings in Experiment 3 do not endorse exorbitant reverberant conditions as solutions to auditory distraction. Nor do they support highly reverberant conditions due to eschewing sound absorption in some office design philosophies, e.g., those with activity-based working (ABW). Instead, Experiment 3's findings support sensible room acoustic and workplace design considerations, as noted previously in other studies [1,3,9,22].

These findings collectively highlight, at the very least, the importance of using representative speech and nonspeech sound conditions when studying the ISE using serial recall tasks in complex acoustic scenarios such as OPOs. Yet, it is worth noting that while the current acoustic presentation is an improvement over studies using simple irrelevant sounds (e.g., words, short sentences, etc.), it is but an instance of a simplified version of actual OPO acoustics. Hence, the current findings are not intended to challenge the basis of studies using simpler irrelevant sounds. Moreover, many other relevant variations of acoustic presentation are possible, and recommended for future studies. If anything, findings here present another piece in the puzzle for understanding cognitive performance in verbal STM tasks, wherein studies with both simple and complex acoustic representation may be justified. An important aspect in the current study was the use of the less common auditory-verbal serial recall task (to represent auditory information in OPOs). However, the STM performance trends were similar to those reported in several ISE studies using the more traditional visual presentation of verbal information. Besides visual-and auditory-verbal serial recall performance has been shown to be similar previously [16].

In general, it is suggested that ISE studies methodically describe and critically examine the context of irrelevant sounds (e.g., sounds used and the reproduction method) much more than is common in general. This should enable more extensive studies that investigate aspects such as ecological validity in a more controlled manner, which is relevant for both simple and complex acoustic presentations. Within this backdrop, it is worth addressing the acoustic conditions in the context of the results, and in terms of limitations of the current experimental design. As mentioned in section III.A.1, SPL presentation across all conditions (except *Quiet*) was kept the same, which is not realistic. In terms of acoustic reproduction, the binaural method here used generic HRTFs. It is possible that the attentional grabbing nature of binaural presentation could vary [92] due to spatial and binaural unmasking effects in more ecologically valid [10] settings including individual/individualized HRTFs. Signal-based changing-state characterization using objective parameters (Tables 1 and 3) has limitations (sections II.C and III.C), but the parameter *FDCC* provided more reliable predictions than *FS*. The finding that *FS* is not a reliable predictor of ISE is consistent with at least one previous study [29], although some studies have suggested otherwise [28,72]. Moreover, the subjective nature of sound design choices within these conditions cannot be overlooked either (section II.C). To address such limitations, a conservative approach is proposed. Herein, given their higher fidelity to actual OPO acoustic environments, it is suggested that the scope of current findings be aligned very closely with the acoustic conditions and other methodological choices herein, and generalizations (e.g., for large OPOs, and different acoustic presentations) be reserved for future research.

In that regard, the following research directions are proposed: considering the ISE (and auditory distraction in general) within variations in 'busyness' and hence changing states of acoustic events for both speech and nonspeech sounds; testing the effect of intonation, emotionality, etc. of voices and having more realistic speech content than current (e.g., using Lombard speech); systematically comparing ISE in quantified changing-state experimental conditions using parameters such as the *FDCC* (and perhaps even *FS*); more acoustically accurate simulation approaches (e.g., SPL reflecting spatial separation of sources, individualized HRTFs, room acoustics for nearby and far reflections, etc.); more realistic presentation of auditory to-be-remembered items and possible masking mechanisms involved for to-be-remembered items due to irrelevant sounds (and vice-versa); using multimodal stimulus presentation and measuring physiological responses during task performance (e.g., [35]); and comparing semantic effects in traditional tasks such as visual-verbal and auditory-verbal short-term memory serial recall with those believed to involve extensive semantic processing (e.g., writing, etc. [36,40,56]); and exploring greater task engagement to study the role of attentional capture [56]. Besides these largely acoustic considerations, it is worth noting that the current experiments were conducted in laboratory conditions with limited correspondence to actual OPOs in terms of visuals, etc. The current laboratory conditions (excluding acoustics) represented those in many such laboratories and test booths worldwide, and hence allow easier repetition by other groups in the future. Regardless, for higher ecological validity, the use of mock offices with realistic visuals, furniture, etc., and/or testing in actual OPOs is highly recommended.

From an even broader perspective, current findings present a case for pushing the boundaries of representative acoustic conditions in laboratory-based experiments. This seems most pertinent for studies like the current that are at the crossroads of multiple disciplines, and the need for such studies to occupy a space alongside studies following a more traditional and more limited acoustic design. Moreover, there is also a possibility of repeated findings from studies with non-representative acoustic design, which, while following established experimental paradigms such as the ISE in serial-recall task, may not follow expectations within realistic settings such as OPOs and similar application areas. If studies such as the current are not conducted and their results are not further explored within basic and applied research, the authors believe that it would stifle the otherwise accelerated and comprehensive growth possible in understanding cognitive phenomenon related to auditory distraction from either end of the ecological validity continuum. This continuum is meant here to span simple laboratory acoustic representations to corresponding real-world instances, and the current findings fit in a region closer to the latter.

CRediT AUTHOR STATEMENT

**Manuj Yadav**: Conceptualization, Data curation, Formal analysis, Investigation, Methodology, Project administration, Resources, Software, Validation, Visualization, Writing – original draft. **Markus Georgi**: Conceptualization, Investigation, Writing - Review & Editing. **Larissa Leist**: Conceptualization, Software, Writing - Review & Editing. **Maria Klatte**:

Conceptualization, Methodology, Supervision, Funding acquisition, Writing - Review & Editing. **Sabine J. Schlittmeier**: Conceptualization, Methodology, Supervision, Funding acquisition, Writing - Review & Editing. **Janina Fels**: Conceptualization, Methodology, Supervision, Funding acquisition, Writing - Review & Editing.

## ACKNOWLEDGMENTS

This study was funded through the Deutsche Forschungsgemeinschaft (DFG, German Research Foundation) Research Grant scheme (Project number: 401278266). The authors wish to thank Sarah Scott, Ulrik Deneken, and Shuyao Guo for help in the experiment design.

DATA AVAILABILITY STATEMENT

The sound conditions and spoken digits used in the experiments here, and the sound conditions in a previous study using classroom sounds [43] are available at https://osf.io/hzjnw/

## REFERENCES

[1] Haapakangas A, Hongisto V, Eerola M, Kuusisto T. Distraction distance and perceived disturbance by noise—An analysis of 21 open-plan offices. The Journal of the Acoustical Society of America 2017;141:127–36. https://doi.org/10.1121/1.4973690.
[2] Pierrette M, Parizet E, Chevret P, Chatillon J. Noise effect on comfort in open-space offices: development of an assessment questionnaire. Ergonomics 2015;58:96–106. https://doi.org/10.1080/00140139.2014.961972.
[3] Yadav M, Cabrera D, Kim J, Fels J, de Dear R. Sound in occupied open-plan offices: Objective metrics with a review of historical perspectives. Applied Acoustics 2021;177:107943. https://doi.org/10.1016/j.apacoust.2021.107943.
[4] Hughes RW, Marsh JE. The functional determinants of short-term memory: Evidence from perceptual-motor interference in verbal serial recall. Journal of Experimental Psychology: Learning, Memory, and Cognition 2017;43:537–51. https://doi.org/10.1037/xlm0000325.
[5] Hughes RW. Auditory distraction: A duplex-mechanism account: Duplex-mechanism account of auditory distraction. PsyCh Journal 2014;3:30–41. https://doi.org/10.1002/pchj.44.
[6] Ellermeier W, Zimmer K. The psychoacoustics of the irrelevant sound effect. Acoustical Science and Technology 2014;35:10–6. https://doi.org/10.1250/ast.35.10.
[7] Colle HA, Welsh A. Acoustic masking in primary memory. Journal of Verbal Learning and Verbal Behavior 1976;15:17–31. https://doi.org/10.1016/S0022-5371(76)90003-7.
[8] Haapakangas A, Hongisto V, Liebl A. The relation between the intelligibility of irrelevant speech and cognitive performance—A revised model based on laboratory studies. Indoor Air 2020;30:1130–46. https://doi.org/10.1111/ina.12726.
[9] ISO 3382-3. ISO 3382-3:2022. Measurement of room acoustic parameters. Part 3: Open plan offices. International Organization for Standardization, Geneva, Switzerland; 2022.
[10] Keidser G, Naylor G, Brungart DS, Caduff A, Campos J, Carlile S, et al. The Quest for Ecological Validity in Hearing Science: What It Is, Why It Matters, and How to Advance It. Ear Hear 2020;41:5S-19S. https://doi.org/10.1097/AUD.0000000000000944.

[11] Haapakangas A, Hongisto V, Hyönä J, Kokko J, Keränen J. Effects of unattended speech on performance and subjective distraction: The role of acoustic design in open-plan offices. Applied Acoustics 2014;86:1–16. https://doi.org/10.1016/j.apacoust.2014.04.018.

[12] Yadav M, Cabrera D. Two simultaneous talkers distract more than one in simulated multi-talker environments, regardless of overall sound levels typical of open-plan offices. Applied Acoustics 2019;148:46–54. https://doi.org/10.1016/j.apacoust.2018.12.007.

[13] Zaglauer M, Drotleff H, Liebl A. Background babble in open-plan offices: A natural masker of disruptive speech? Applied Acoustics 2017;118:1–7. https://doi.org/10.1016/j.apacoust.2016.11.004.

[14] Hanley JR, Bourgaize J. Similarities between the irrelevant sound effect and the suffix effect. Mem Cogn 2018;46:841–8. https://doi.org/10.3758/s13421-018-0806-8.

[15] Klatte M, Lee N, Hellbruck J. Effects of irrelevant speech and articulatory suppression on serial recall of heard and read materials. Psychologische Beiträge 2002;44:166.

[16] Schlittmeier SJ, Hellbrück J, Klatte M. Does irrelevant music cause an irrelevant sound effect for auditory items? European Journal of Cognitive Psychology 2008;20:252–71. https://doi.org/10.1080/09541440701427838.

[17] Yadav M, Kim J, Cabrera D, de Dear R. Auditory distraction in open-plan office environments: The effect of multi-talker acoustics. Applied Acoustics 2017;126:68–80. https://doi.org/10.1016/j.apacoust.2017.05.011.

[18] Jones DM, Macken WJ, Murray AC. Disruption of visual short-term memory by changing-state auditory stimuli: The role of segmentation. Memory & Cognition 1993;21:318–28. https://doi.org/10.3758/BF03208264.

[19] Bell R, Mieth L, Röer JP, Buchner A. The metacognition of auditory distraction: Judgments about the effects of deviating and changing auditory distractors on cognitive performance. Mem Cogn 2022;50:160–73. https://doi.org/10.3758/s13421-021-01200-2.

[20] Perham N, Hodgetts H, Banbury S. Mental arithmetic and non-speech office noise: An exploration of interference-by-content. Noise and Health 2013;15:73. https://doi.org/10.4103/1463-1741.107160.

[21] Yadav M, Cabrera D, Kim J, Love J, Holmes J, Fels J, et al. Investigating noise disturbance in open-plan offices using measurements of the room acoustics, and of the sound environment during occupancy. Proceedings of Euronoise 2021, Madeira, Portugal: 2021.

[22] ISO 22955. ISO 22955 Acoustics — Acoustic quality of open office spaces 2021.

[23] Buchner A, Rothermund K, Wentura D, Mehl B. Valence of distractor words increases the effects of irrelevant speech on serial recall. Memory & Cognition 2004;32:722–31.

[24] LeCompte DC, Neely CB, Wilson JR. Irrelevant speech and irrelevant tones: The relative importance of speech to the irrelevant speech effect. Journal of Experimental Psychology: Learning, Memory, and Cognition 1997;23:472–83. https://doi.org/10.1037/0278-7393.23.2.472.

[25] Röer JP, Bell R, Körner U, Buchner A. A semantic mismatch effect on serial recall: Evidence for interlexical processing of irrelevant speech. Journal of Experimental Psychology: Learning, Memory, and Cognition 2019;45:515–25. https://doi.org/10.1037/xlm0000596.

[26] Röer JP, Bell R, Buchner A, Saint-Aubin J, Sonier R-P, Marsh JE, et al. A multilingual preregistered replication of the semantic mismatch effect on serial recall. Journal of

Experimental Psychology: Learning, Memory, and Cognition 2022;48:966–74. https://doi.org/10.1037/xlm0001066.

[27] Marsh JE, Hughes RW, Jones DM. Interference by process, not content, determines semantic auditory distraction. Cognition 2009;110:23–38. https://doi.org/10.1016/j.cognition.2008.08.003.

[28] Schlittmeier SJ, Weißgerber T, Kerber S, Fastl H, Hellbrück J. Algorithmic modeling of the irrelevant sound effect (ISE) by the hearing sensation fluctuation strength. Attention, Perception, & Psychophysics 2012;74:194–203. https://doi.org/10.3758/s13414-011-0230-7.

[29] Ellermeier W, Kattner F, Ueda K, Doumoto K, Nakajima Y. Memory disruption by irrelevant noise-vocoded speech: Effects of native language and the number of frequency bands. The Journal of the Acoustical Society of America 2015;138:1561–9. https://doi.org/10.1121/1.4928954.

[30] Wöstmann M, Obleser J. Acoustic Detail But Not Predictability of Task-Irrelevant Speech Disrupts Working Memory. Frontiers in Human Neuroscience 2016;10.

[31] Vachon F, Marsh JE, Labonté K. The automaticity of semantic processing revisited: Auditory distraction by a categorical deviation. Journal of Experimental Psychology: General 2020;149:1360–97. https://doi.org/10.1037/xge0000714.

[32] Labonté K, Marsh JE, Vachon F. Distraction by Auditory Categorical Deviations Is Unrelated to Working Memory Capacity: Further Evidence of a Distinction between Acoustic and Categorical Deviation Effects. Auditory Perception & Cognition 2022;0:1–26. https://doi.org/10.1080/25742442.2022.2033109.

[33] Littlefair Z, Vachon F, Ball LJ, Robinson N, Marsh JE. Acoustic, and Categorical, Deviation Effects are Produced by Different Mechanisms: Evidence from Additivity and Habituation. Auditory Perception & Cognition 2022;5:1–24. https://doi.org/10.1080/25742442.2022.2063609.

[34] Röer JP, Buchner A, Bell R. Auditory Distraction in Short-term Memory: Stable Effects of Semantic Mismatches on Serial Recall. Auditory Perception & Cognition 2019;2:143–62. https://doi.org/10.1080/25742442.2020.1722560.

[35] Ríos-López P, Widmann A, Bidet-Caulet A, Wetzel N. The effect of background speech on attentive sound processing: A pupil dilation study. International Journal of Psychophysiology 2022;174:47–56. https://doi.org/10.1016/j.ijpsycho.2022.02.003.

[36] Schlittmeier SJ, Marsh JE. Review of research on the effects of noise on cognitive performance 2017-2021, Stockholm: 2021.

[37] Roberts W, Milich R, Fillmore MarkT. The Effects of Preresponse Cues on Inhibitory Control and Response Time in Adults With ADHD. Journal of Attention Disorders 2013;20:317–24. https://doi.org/10.1177/1087054713495737.

[38] Schlittmeier SJ, Hellbrück J. Background music as noise abatement in open-plan offices: A laboratory study on performance effects and subjective preferences. Applied Cognitive Psychology 2009;23:684–97. https://doi.org/10.1002/acp.1498.

[39] Jones D, Macken W. Auditory Babble and Cognitive Efficiency: Role of Number of Voices and Their Location. J Exp Psychol Appl 1995;1:216–26.

[40] Keus van de Poll M, Carlsson J, Marsh JE, Ljung R, Odelius J, Schlittmeier SJ, et al. Unmasking the effects of masking on performance: The potential of multiple-voice

masking in the office environment. The Journal of the Acoustical Society of America 2015;138:807–16. https://doi.org/10.1121/1.4926904.
[41] Buchner A, Bell R, Rothermund K, Wentura D. Sound source location modulates the irrelevant-sound effect. Memory & Cognition 2008;36:617–28. https://doi.org/10.3758/MC.36.3.617.
[42] Schlittmeier SJ, Hellbrück J, Thaden R, Vorländer M. The impact of background speech varying in intelligibility: Effects on cognitive performance and perceived disturbance. Ergonomics 2008;51:719–36. https://doi.org/10.1080/00140130701745925.
[43] Leist L, Breuer C, Yadav M, Fremerey S, Fels J, Raake A, et al. Differential Effects of Task-Irrelevant Monaural and Binaural Classroom Scenarios on Children's and Adults' Speech Perception, Listening Comprehension, and Visual–Verbal Short-Term Memory. International Journal of Environmental Research and Public Health 2022;19:15998. https://doi.org/10.3390/ijerph192315998.
[44] Bronkhorst AW. The cocktail-party problem revisited: early processing and selection of multi-talker speech. Attention, Perception, & Psychophysics 2015;77:1465–87. https://doi.org/10.3758/s13414-015-0882-9.
[45] Culling JF, Lavandier M. Binaural Unmasking and Spatial Release from Masking. In: Litovsky RY, Goupell MJ, Fay RR, Popper AN, editors. Binaural Hearing, vol. 73, Cham: Springer International Publishing; 2021, p. 209–41. https://doi.org/10.1007/978-3-030-57100-9_8.
[46] Shinn-Cunningham BG. Object-based auditory and visual attention. Trends in Cognitive Sciences 2008;12:182–6. https://doi.org/10.1016/j.tics.2008.02.003.
[47] Westermann A, Buchholz JM. The influence of informational masking in reverberant, multi-talker environments. The Journal of the Acoustical Society of America 2015;138:584–93. https://doi.org/10.1121/1.4923449.
[48] Westermann A, Buchholz JM. The effect of nearby maskers on speech intelligibility in reverberant, multi-talker environmentsa). The Journal of the Acoustical Society of America 2017;141:2214–23. https://doi.org/10.1121/1.4979000.
[49] Reinten J, Braat-Eggen PE, Hornikx M, Kort HSM, Kohlrausch A. The indoor sound environment and human task performance: A literature review on the role of room acoustics. Building and Environment 2017;123:315–32. https://doi.org/10.1016/j.buildenv.2017.07.005.
[50] Beaman CP, Holt NJ. Reverberant auditory environments: the effects of multiple echoes on distraction by 'irrelevant' speech. Applied Cognitive Psychology 2007;21:1077–90. https://doi.org/10.1002/acp.1315.
[51] Perham N, Banbury S, Jones DM. Do realistic reverberation levels reduce auditory distraction? Applied Cognitive Psychology 2007;21:839–47. https://doi.org/10.1002/acp.1300.
[52] Vachon F, Winder E, Lavandier M, Hughes RW. The bigger the better and the more the merrier? Realistic office reverberation levels abolish cognitive distraction by multiple-voice speech. 12th ICBEN Congress on Noise as a Public Health Problem, Zurich: 2017, p. 10.
[53] Arweiler I, Buchholz JM. The influence of spectral characteristics of early reflections on speech intelligibility. The Journal of the Acoustical Society of America 2011;130:996–1005. https://doi.org/10.1121/1.3609258.

[54] Bradley JS, Sato H, Picard M. On the importance of early reflections for speech in rooms. The Journal of the Acoustical Society of America 2003;113:3233. https://doi.org/10.1121/1.1570439.

[55] Marsh JE, Yang J, Qualter P, Richardson C, Perham N, Vachon F, et al. Postcategorical auditory distraction in short-term memory: Insights from increased task load and task type. Journal of Experimental Psychology: Learning, Memory, and Cognition 2018;44:882–97. https://doi.org/10.1037/xlm0000492.

[56] Marsh JE, Ljung R, Jahncke H, MacCutcheon D, Pausch F, Ball LJ, et al. Why are background telephone conversations distracting? Journal of Experimental Psychology: Applied 2018;24:222–35. https://doi.org/10.1037/xap0000170.

[57] Scott SK, Rosen S, Beaman CP, Davis JP, Wise RJS. The neural processing of masked speech: Evidence for different mechanisms in the left and right temporal lobes. The Journal of the Acoustical Society of America 2009;125:1737–43. https://doi.org/10.1121/1.3050255.

[58] Senan TU, Jelfs S, Kohlrausch A. Cognitive disruption by noise-vocoded speech stimuli: Effects of spectral variation. The Journal of the Acoustical Society of America 2018;143:1407–16. https://doi.org/10.1121/1.5026619.

[59] Blazier WE. Revised noise criteria for application in the acoustical design and rating of HVAC systems. Noise Control Eng 1981;16:64–73.

[60] Yadav M, Cabrera D, Love J, Kim J, Holmes J, Caldwell H, et al. Reliability and repeatability of ISO 3382-3 metrics based on repeated acoustic measurements in open-plan offices. Applied Acoustics 2019;150:138–46. https://doi.org/10.1016/j.apacoust.2019.02.010.

[61] Veitch JA, Bradley JS, Legault LM, Norcross S, Svec JM. Masking speech in open-plan offices with simulated ventilation noise: noise level and spectral composition effects on acoustic satisfaction. Institute for Research in Construction, Internal Report IRC-IR-846 2002.

[62] Monk A, Carroll J, Parker S, Blythe M. Why are mobile phones annoying? Behaviour & Information Technology 2004;23:33–41. https://doi.org/10.1080/01449290310001638496.

[63] Monk A, Fellas E, Ley E. Hearing only one side of normal and mobile phone conversations. Behaviour & Information Technology 2004;23:301–5. https://doi.org/10.1080/01449290410001712744.

[64] Emberson LL, Lupyan G, Goldstein MH, Spivey MJ. Overheard Cell-Phone Conversations: When Less Speech Is More Distracting. Psychological Science 2010;21:1383–8. https://doi.org/10.1177/0956797610382126.

[65] Galván VV, Vessal RS, Golley MT. The Effects of Cell Phone Conversations on the Attention and Memory of Bystanders. PLOS ONE 2013;8:e58579. https://doi.org/10.1371/journal.pone.0058579.

[66] ISO 7250-3. Basic human body measurements for technological design Part 3: Worldwide and regional design ranges for use in product standards. International Organization for Standardization, Geneva, Switzerland; 2015.

[67] Schröder D. Physically based real-time auralization of interactive virtual environments. vol. 11. Logos Verlag Berlin GmbH; 2011.

[68] Brinkmann F, Lindau A, Weinzierl S, Par S van de, Müller-Trapet M, Opdam R, et al. A High Resolution and Full-Spherical Head-Related Transfer Function Database for Different Head-Above-Torso Orientations. JAES 2017;65:841–8.

[69] Oberem J, Fels J. Speech Material for a Paradigm on the Intentional Switching of Auditory Selective Attention 2020. https://doi.org/10.18154/RWTH-2020-02105.

[70] IEC 60318-1. Simulators of human head and ear — Part 1: Ear simulator for the calibration of supra-aural earphones. International Electrotechnical Commission, Geneva, Switzerland; 1998.

[71] Fastl H, Zwicker E. Psychoacoustics: Facts and models. Berlin, Heidelberg: Springer Berlin Heidelberg; 2007. https://doi.org/10.1007/978-3-540-68888-4.

[72] Kattner F, Ellermeier W. Emotional prosody of task-irrelevant speech interferes with the retention of serial order. Journal of Experimental Psychology: Human Perception and Performance 2018;44:1303–12. https://doi.org/10.1037/xhp0000537.

[73] Park M, Kohlrausch A, van Leest A. Irrelevant speech effect under stationary and adaptive masking conditions. The Journal of the Acoustical Society of America 2013;134:1970. https://doi.org/10.1121/1.4816939.

[74] Senan TU. An evaluation of a psychoacoustic model of the changing-state hypothesis. PhD Thesis. Technische Universiteit Eindhoven, 2019.

[75] Zhou T, Zhang M-J, Li C. A Model for Calculating Psychoacoustical Fluctuation Strength. Journal of the Audio Engineering Society 2015;63:713–24. https://doi.org/10.17743/jaes.2015.0070.

[76] Cabrera D, Jimenez D, Martens WL. Audio and Acoustical Response Analysis Environment (AARAE): a tool to support education and research in acoustics. Proceedings of Internoise, Melbourne, Australia: 2014.

[77] Braat-Eggen E, Reinten J, Hornikx M, Kohlrausch A. The influence of background speech on a writing task in an open-plan study environment. Building and Environment 2020;169:106586. https://doi.org/10.1016/j.buildenv.2019.106586.

[78] Gelman A, Carlin JB, Stern HS, Rubin DB. Bayesian data analysis (Vol. 2). Taylor & Francis Boca Raton; 2014.

[79] Kruschke J. Doing Bayesian data analysis: A tutorial with R, JAGS, and Stan. Academic Press; 2014.

[80] Bürkner P-C. brms: An R Package for Bayesian Multilevel Models Using Stan. Journal of Statistical Software 2017;80:1–28. https://doi.org/10.18637/jss.v080.i01.

[81] Makowski D, Ben-Shachar MS, Chen SHA, Lüdecke D. Indices of Effect Existence and Significance in the Bayesian Framework. Frontiers in Psychology 2019;10.

[82] Makowski D, Ben-Shachar M, Lüdecke D. bayestestR: Describing Effects and their Uncertainty, Existence and Significance within the Bayesian Framework. JOSS 2019;4:1541. https://doi.org/10.21105/joss.01541.

[83] Alikadic L, Röer JP. Loud Auditory Distractors Are More Difficult to Ignore After All. Experimental Psychology 2022;69:163–71. https://doi.org/10.1027/1618-3169/a000554.

[84] IEC 60318-16 Ed.5. Sound system equipment - Part 16: Objective rating of speech intelligibility by speech transmission index. International Electrotechnical Commission, Geneva, Switzerland; 2020.

[85] Berzborn M, Bomhardt R, Klein J, Richter J-G, Vorländer M. The ITA-Toolbox: An Open Source MATLAB Toolbox for Acoustic Measurements and Signal Processing, 43th Annual German Congress on Acoustics, Kiel (Germany), 6 Mar 2017 - 9 Mar 2017; 2017.

[86] Bradley JS. The acoustical design of conventional open plan offices. Canadian Acoustics 2003;31:23–31.
[87] Virjonen P, Keränen J, Hongisto V. Determination of Acoustical Conditions in Open-Plan Offices: Proposal for New Measurement Method and Target Values. Acta Acustica United with Acustica 2009;95:279–90. https://doi.org/10.3813/AAA.918150.
[88] Cooke M, Lu Y. Spectral and temporal changes to speech produced in the presence of energetic and informational maskersa). The Journal of the Acoustical Society of America 2010;128:2059. https://doi.org/10.1121/1.3478775.
[89] Bottalico P, Graetzer S, Hunter EJ. Effects of speech style, room acoustics, and vocal fatigue on vocal effort. The Journal of the Acoustical Society of America 2016;139:2870–9. https://doi.org/10.1121/1.4950812.
[90] Colle HA. Auditory encoding in visual short-term recall: effects of noise intensity and spatial location. Journal of Verbal Learning and Verbal Behavior 1980;19:722–35. https://doi.org/10.1016/S0022-5371(80)90403-X.
[91] Ellermeier W, Hellbrück J. Is level irrelevant in "irrelevant speech"? Effects of loudness, signal-to-noise ratio, and binaural unmasking. Journal of Experimental Psychology: Human Perception and Performance 1998;24:1406–14. https://doi.org/10.1037/0096-1523.24.5.1406.
[92] Oberem J, Lawo V, Koch I, Fels J. Intentional Switching in Auditory Selective Attention: Exploring Different Binaural Reproduction Methods in an Anechoic Chamber. Acta Acustica United with Acustica 2014;100:1139–48. https://doi.org/10.3813/AAA.918793.